\documentclass{ws-mpla}

\begin{document}

\markboth{Michele Trapletti}{Gauge symmetry breaking in
orbifold model building}

\catchline{}{}{}{}{}

\title{GAUGE SYMMETRY BREAKING IN ORBIFOLD MODEL
BUILDING}

\author{\footnotesize MICHELE TRAPLETTI}

\address{
Institut f\"ur Theoretische Physik,
Ruprecht-Karls-Universit\"at Heidelberg,\\
Philosophenweg 16, D-69120, Heidelberg
Germany.\\ m.trapletti@thphys.uni-heidelberg.de}

\maketitle


\begin{abstract}
We review the gauge symmetry breaking mechanism due to orbifold
projections in orbifold model building.
We explicitly show the existence of a scale of breaking if such
a symmetry breaking is due to freely-acting orbifold operators only,
i.e. in case the breaking is realized {\it non-locally} in the
internal space.
We show that such a scale is related to the compactification moduli
only, and that there are no extra continuous parameters, at least in
semirealistic models with $\mathcal N=1$ SUSY in four dimensions.
In this sense, the mechanism is peculiarly different from the standard
Higgs (or Hosotani) symmetry breaking mechanism.
We show that the mechanism also differs from that present in standard
orbifold models where, even in presence of discrete Wilson lines,
a scale of breaking is generically missing, since the breaking
is {\it localized} in specific points in the internal space.

We review a set of background geometries where the described
{\it non-local} breaking is realized, both in the case of two
and six extra dimensions. 
In the latter case, relevant in string model building, we consider
both heterotic and open string compactifications.
\end{abstract}
\ccode{PACS Nos.: 11.10.Kk, 11.25.Mj, 11.25.Wx, 12.10.-g}
\ccode{Preprint No: HD-THEP-06-27}

\section{Introduction and outline}

Unification of Electroweak and Strong interaction couplings\cite{LateRunning},
at a precise energy scale\cite{Amaldi:1991cn}, is the main information
we have about physics at energies much larger than ${\rm TeV}$.
Given how elementary particles fit multiplets of a unified gauge
group, it is natural to conclude that such unification is not an
accident, but a crucial feature of Nature, that should be explained,
rather then merely reproduced, by an extension of the Standard Model
of Particle Physics.

Many attempts have been devoted to the construction of unified models,
in a field theory\cite{LateUnif} scenario, even in presence
of one\cite{orbgutfield5,orbgutfield5b} or two\cite{orbgutfield6}
extra dimensions, as well as in string model building 
(for recent realistic constructions see\cite{orbgutstring}).
In the latter, large gauge groups naturally arise, and a unified 
description of gauge and gravity interactions is also possible.
In all the models a key feature is the gauge symmetry breaking
mechanism, and in this paper we focus on such an issue:
Our main aim is the study of the mechanism due to the presence of 
orbifold projections in (toroidal) orbifold models.
We avoid a complete discussion of other symmetry breaking mechanisms,
such as the one realized by continuous Wilson lines or by Higgs-like
effects, as well as the kind of gauge symmetry breaking due to magnetic
fluxes for the gauge fields or, in a dual picture, the one due to the
presence of angles between D-branes in open string model building.

We distinguish between two different options, characterized by the 
details of the orbifold projection. Namely, it may act in the internal
space as a {\it rotation} (with fixed points) or as a
{\it rototranslation} (without fixed points i.e. {\it freely}).

In the {\it rotational} case the breaking is {\it localized} in the
fixed points and, generically, there is no scale of breaking: the coupling
constants of the unbroken gauge factors (namely, $SU(3)$, $SU(2)$
and $U(1)$) run in different ways at all the energies. This implies
that the unification of the couplings can be at best numerically reproduced, 
but not explained.

In the {\it rototranslational} case the breaking is 
{\it non-local}\mbox{\hspace{1pt}}\cite{Hosotani:1983xw,SS}\footnote{Notice that
such a construction is also a realization of the well known Scherk-Schwarz
(SUSY) breaking mechanism\cite{SS}, as shown, in a string context, 
in\cite{free,Antoniadis:1999ux}.  (For similar constructions in a field
theory context, see also\cite{SSf}.)}.
This implies that the couplings run in the same unified way between
the high energy cutoff of the model (string scale) and an intermediate
scale, related to the compactification scales.
Then, from this scale down to low energy, the three couplings run in
different ways, and so they split apart from each other\cite{Hall:2001xb}.
It is very natural to define such an intermediate scale as the
symmetry breaking scale of the model.

The presence of such a scale is crucial in heterotic string model building, 
where the string scale is typically larger than the ``observed'' unification
scale $M_{GUT}=3\times 10^{16}\,{\rm GeV}$\cite{Kaplunovsky:1985yy}: the
hierarchy $M_{string}/M_{GUT}$ can be introduced via a compactification
scale smaller then $M_{string}$, rather then by appealing to large
threshold corrections at the string
scale\cite{Dixon:1990pc}\footnote{For more recent alternatives addressing
the $M_{string}/M_{GUT}$ hierarchy see also\cite{Faraggi:1993sr}.}.

We mainly focus on non-local breaking, describing its details in
nontrivial compactifications.
Since an orbifold model with a single orbifold operator/projection is
an over-simplified system\footnote{Indeed, a toroidal orbifold with a
single orbifold operator, acting freely, is typically inconsistent with
the requirement of having a four-dimensional chiral spectrum.},
we extend the analysis
to orbifold models with enlarged orbifold group. In this case
the gauge symmetry breaking is non-local only provided that all the
gauge-symmetry-breaking orbifold operators have a free
action\cite{Hebecker:2003we}.
Such a requirement imposes severe constraints on the allowed 
orbifolds/geometries; we review those described 
in\cite{Hebecker:2004ce,Trapletti:2005ey}, relevant in string model
building.
The complication of having many orbifold operators has a relevant extra
consequence. Namely, it ensures the absence of continues Wilson lines
mimicking the kind of gauge symmetry breaking we are interested in,
i.e. it ensures that the symmetry breaking scale is completely fixed by
the compactification moduli.
The mechanism is then peculiarly different from the Hosotani mechanism,
where an extra continuous parameter, other than the compactification scale,
can be tuned to fix the breaking scale, exactly as in the standard Higgs
mechanism.
In this sense, the mechanism combines the absence of extra continuous
parameters, typical of discrete Wilson lines breaking, with the presence of
a breaking scale, typical of Hosotani/Higgs mechanism.

The Outline of the paper is the following\\
In Sect.~2 we consider field theory examples, in presence of two 
extra dimensions compactified on orbifolds.
We first study two simplified models, one with rotational action, the other
with rototranslational action, computing the running of the coupling
constants. We show  that only in the rototranslational case it is possible
to have unification of the couplings at the compactification scale and
unified running of the couplings between this scale and the cutoff.
We extend the study to a third model with non-local breaking, discussed
in\cite{Hebecker:2003we} and relevant also in\cite{Hebecker:2004ce}.
We compute the symmetry breaking scale, finding its exact relation with
the surface of the two-dimensional space. We also show the absence of
continuous Wilson lines mimicking the non-local gauge symmetry breaking. 
Finally, we extend the results to generic orbifolds, and comment about the
gauge symmetry breaking  due to discrete Wilson lines. We show that in
the latter case a scale of breaking is present only in {\it very specific}
constructions and is absent in the general case.

In Sect.~3 we study 6d orbifold geometries, relevant in string model
building since Type I and heterotic string compactifications on such
backgrounds produce 4d $\mathcal N=1$ SUSY models with non-local
gauge symmetry breaking.
We comment about the details of open string model
building\cite{Trapletti:2005ey}. We remind that, in presence of stacks
of Dp-branes with $p<9$, the orbifold action can be {\it external} to
the stacks, i.e. it can identify different stacks of D-branes. In this case the
symmetry breaking reduces the rank of the gauge group.
We comment about a purely external orbifold action, irrelevant for
our purposes, a purely internal action, producing models with features
close to the heterotic ones, and finally we address the issue of a mixed
internal and external action. We show that in this case the combination
of a rank-reducing orbifold with a discrete Wilson line can produce a model
with gauge coupling unification at a scale fixed by the compactification scale.

\begin{center}
{\bf Acknowledgments}
\end{center}
\mbox{}
\vspace{-18pt}

\noindent
It is a pleasure to thank Arthur Hebecker and Stefan Groot Nibbelink
for discussions and comments on the draft.

\section{Gauge symmetry breaking via orbifold projections}
Our starting point is $SU(5)$ gauge theory defined on a $4+2$ dimensional
space-time.
The two extra dimensions are compactified on a toroidal orbifold.
Such a space is a 2d torus with complex parameter $z=z_1+i z_2$
having periodicities $z_i\sim z_i+2\pi R_i$, in which
we identify points  that are symmetric under the action of the orbifold
group $\mathcal G$: $z\equiv g(z)\,\forall\, g\in\mathcal G$.
We consider bulk fields periodic in the internal dimensions,
allowing non-trivial orbifold identifications, i.e. we consider a 
Kaluza-Klein expansion reduced to the states that are invariant under
the orbifold action.

As an example, consider a scalar field in the fundamental representation of
$SU(5)$, $\Phi^a(x^\mu,z)$. It has Kaluza-Klein (KK) expansion
\begin{equation}
\Phi^a (x^\mu,z)=\sum_{m_i=-\infty}^\infty \Phi^a_{m_1,m_2}(x^\mu) 
e^{\frac{i m_1 z_1}{R_1}+\frac{i m_2 z_2}{R_2}}.
\end{equation}
The action of the orbifold operator $g$ is
\begin{equation}
\label{exp}
g:\,\Phi^a(x^\mu,z)=\gamma_g^{ab} \Phi^b(x^\mu,g(z)),
\end{equation}
with $\gamma_g$ the embedding of the orbifold group in the
gauge bundle. We project out of the spectrum all the KK states that are
not invariant under the action of $g$.
Considering the $Z_2$ case $g(z)=-z$, we deduce the action of $g$ on the 
KK modes, and see that only the combinations
\begin{equation}
\label{inv}
\Phi^a_{m_1,m_2}+\gamma_g^{ab} \Phi^b_{-m_1,-m_2}
\end{equation}
are left in the spectrum. We also deduce that, depending on $\gamma_g$, the 
orbifold identification breaks the gauge
symmetry\cite{orbgutfield5,orbgutfield5b}, e.g.
$SU(5)\rightarrow SU(3)\times SU(2)\times U(1)$.

In case an orbifold operator $g$ has fixed points, as in the $Z_2$ example
given above, these result in singularities in the orbifold space itself,
where extra states can be localized.
These states arise, in a string model, due to consistency conditions
(e.g. modular invariance in heterotic string model building),
and must respect the gauge symmetry that is left unbroken in the
singularities, i.e. the gauge symmetry preserved by the operator $g$.

\begin{figure}[t]
\begin{center}
\includegraphics[scale=0.5]{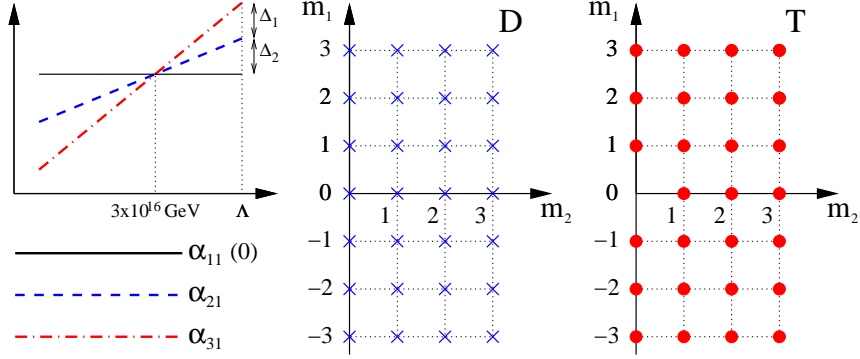}\vspace{-.1cm}
\caption{\footnotesize \it 
Kaluza-Klein expansion of a fundamental representation
of $SU(5)$ in the $T^2/Z_2$ orbifold. The figure in the
middle shows the KK tower of the doublet (D), a blue cross in
a point with coordinates $m_1,m_2$ indicates the presence
in the tower of a single state with KK mass 
$m^2=m_1^2/R_1^2+m_2^2/R_2^2$.
The figure on the right side shows the tower for the triplet (T).
A similar expansion holds for any $SU(5)$ multiplets, this
implies that the running of the difference of the inverse
coupling constants is as given in the figure on the left side,
i.e. it never stops up to the cutoff energy $\Lambda$.
Thus, it can be consistent with low energy data either in case
$\Lambda\sim 10^{16} {\rm GeV}$ or in the presence of threshold
corrections $\Delta_1$ and $\Delta_2$ at the cutoff energy.
In both cases the unification of the couplings at 
$3\times 10^{16} {\rm GeV}$ is nothing else
than a numerical accident.
}\label{kk1}
\end{center}
\end{figure}

We can now study the details of the gauge symmetry breaking, in particular
the existence of an energy scale. To inspect it we compute the 
{\it differential running}\cite{orbgutfield5b} of the
$SU(3)$, $SU(2)$ and $U(1)$ couplings, i.e. the running of the difference
of the inverse couplings: $\alpha_{ij}=\alpha_i^{-1}-\alpha_j^{-1}$.
Such a quantity can be written, in presence of two extra dimensions, as
\begin{eqnarray}
\label{run}
&
\alpha_{ij}(M_Z)=&b_{ij;\,00}
\log\left(\frac{\Lambda}{M_Z}\right)+
\hspace{-12pt}
\sum_{(m_1 m_2)\neq (00)}\hspace{-12pt}
b_{ij;\,m_1 m_2} \log\left(\frac{\Lambda}{\sqrt{m_1^2 R_1^{-2}+
m_2^2 R_2^{-2}}}\right).
\end{eqnarray}
In Eq.~(\ref{run}) $M_Z$ is a generic energy scale, lower than any
compactification scale, while $\Lambda$ is the high energy cutoff of
the model; the coefficients $b_{ij;\,mn}$ are defined 
as $b_{ij;\,mn}=b_{i;\,mn}-b_{j;\,mn}$, with $b_{i;\,mn}$ the standard
$\beta$-function coefficients of the coupling $\alpha_i$, due to KK modes
with mass $M^2=m_1^2/R_1^2+m_2^2/R_2^2$.

If there exists an energy scale $M_{GUT}<\Lambda$, such that for energies
larger than $M_{GUT}$ the contribution of massless modes to $\alpha_{ij}$
is precisely canceled by that of massive modes, than 
Eq.~(\ref{run}) can be written as
\begin{eqnarray}
\label{runc}
\alpha_{ij}(M_Z)=b_{ij;\,00}
\log\left(\frac{M_{GUT}}{M_Z}\right),
\end{eqnarray}
and we identify $M_{GUT}$ with the unification scale. Indeed, from
Eq.~(\ref{runc}) we deduce that the couplings run in a unified
way from $\Lambda$ to $M_{GUT}$, and they split (unify) precisely at
$M_{GUT}$. 
If such an $M_{GUT}$ is not present, then the gauge symmetry breaking
occurs in the absence of a breaking scale.

\subsection{{\it Rotational} orbifold: a $T^2/Z_2$ model with
localized gauge symmetry breaking}
\label{roto}
As an example of rotational orbifold, we consider $SU(5)$ gauge theory
on $T^2/Z_2$.
We take $Z_2=\{I,\,g\}$, $g:\,z\rightarrow -z$. Given the generators
of $SU(5)$ as $5\times 5$ matrices, and choosing $\gamma_g$ as the diagonal
matrix $\gamma_g={\rm Diag}(-1,-1,-1,1,1)$, we have the desired gauge
symmetry breaking 
$SU(5)\rightarrow SU(3)\times SU(2)\times U(1)$\cite{orbgutfield5,orbgutfield5b}.

In order to study the differential running, we need the Kaluza-Klein
expansion of the bulk fields. Consider, as an example, a scalar in the
fundamental of $SU(5)$. It has KK expansion given in Eq.~(\ref{exp}),
with invariant states given by Eq.~(\ref{inv}).
In detail, for each $a=1,\dots,\, 5$, there is a tower of states with 4d mass 
$M^2=m_1^2/R_1^2+m_2^2/R_2^2$. The $g$-projection reduces these towers:
for $a=4,\,5$ (i.e. for the fields forming a doublet of the unbroken $SU(2)$)
there is a tower of states with $m_2\ge 0$,
$m_1\in\mathbb Z$. For $a=1,\,2,\,3$ (i.e. for the fields forming a 
triplet of the unbroken $SU(3)$), there is the same
tower as for the doublet, but with missing massless mode
($m_1=0,\,m_2=0$), as shown in Fig.~\ref{kk1}.
At each massive KK level there is a single surviving doublet and a single
surviving triplet, that combine and form a complete multiplet of $SU(5)$. 
Such a statement holds in the expansion of any multiplet of $SU(5)$:
it splits into submultiplets having identical {\it massive} KK tower,
so that at any non-zero KK level they recombine in a full $SU(5)$
multiplet.
In this way the contribution to the differential running 
due to massive KK modes is just zero, since it is the contribution of
full multiplets of $SU(5)$ only, and, in Eq.~(\ref{run}),
$b_{ij;\,m_1 m_2}=0$ for each $i,j$, provided that $(m_1,m_2)\neq (0,0)$.

Since the operator $g$ has fixed points, the model contains
extra localized matter, that propagates in four dimensions only.
This implies that its  contribution to the differential
running affects $b_{ij;\,00}$ only. Thus, we conclude that
\begin{equation}
\alpha_{ij}(M_Z)=b_{ij;\,00}
\log\left(\frac{\Lambda}{M_Z}\right),
\end{equation}
the differential running of the coupling constants is unaffected by threshold 
corrections (it is the same at all the energy scales), and there is no scale
of breaking.

In the described model, gauge unification is
at best numerically achieved, rather then explained, by either identifying
the cutoff scale with the ``observed'' unification scale 
$3\times \,10^{16}\,{\rm GeV}$, or by introducing threshold
corrections at the cutoff scale, as shown in Fig.~\ref{kk1}.

\begin{figure}[t]
\begin{center}
\includegraphics[scale=0.5]{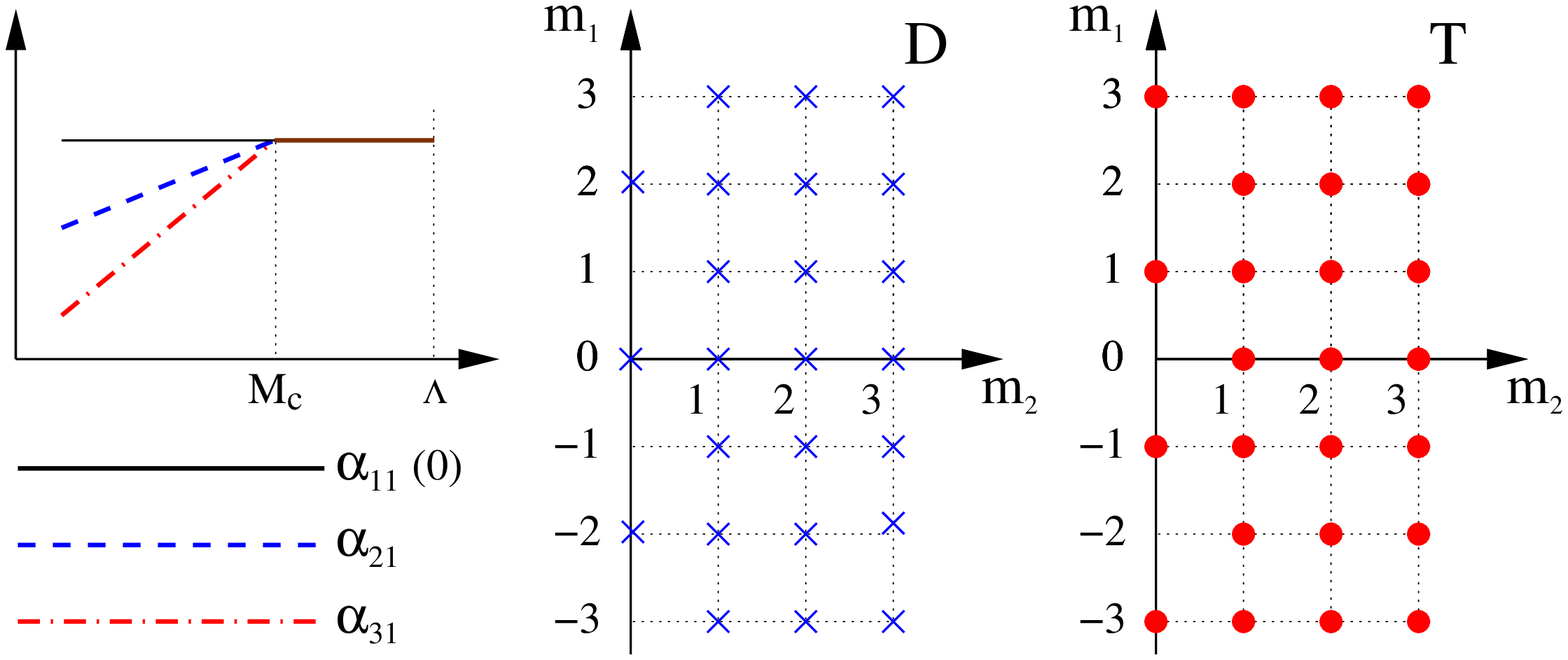}\vspace{-.1cm}
\caption{\footnotesize \it 
Kaluza-Klein expansion of a fundamental representation
of $SU(5)$ in the $T^2/Z_2^\prime$ orbifold. The
figure in the middle shows the KK tower of the doublet (D),
a blue cross in a point with coordinates $m_1,m_2$ indicates
the presence  in the tower of a single state with KK mass 
$m^2=m_1^2/R_1^2+m_2^2/R_2^2$. 
The figure on the right shows the tower for the triplet (T).
A similar expansion holds for any $SU(5)$ multiplets, this
implies that the running of the difference of the inverse
coupling constants is as given in the figure on the left, 
i.e. it precisely stops, due to the KK contributions, 
at the compactification scale $M_C=1/R_1$.
In the model the breaking scale is then reproduced
and explained provided $M_C=M_{GUT}=3\times 10^{16} {\rm GeV}$.
}\label{kk2}
\end{center}
\end{figure}

\subsection{{\it Rototranslational} orbifold: a $T^2/Z_2^\prime$
model with non-local gauge symmetry breaking}
\label{trasl}
We can consider the same $SU(5)$ theory on a different $T^2/Z_2^\prime$
orbifold, with $Z_2^\prime$ generated by $g^\prime$ with geometric action
\begin{equation}
g^\prime:z_1\rightarrow z_1+\pi R_1,\,\,\,\,
g^\prime:z_2\rightarrow -z_2,
\end{equation}
and gauge embedding $\gamma_{g^\prime}={\rm Diag}(-1,-1,-1,1,1)$.
Due to the translational
action along $z_1$, $g^\prime$ acts freely, and the resulting orbifold 
has no singularities.
Consider now the Kaluza-Klein expansion of a scalar field in the
fundamental representation of $SU(5)$, given in Eq.~(\ref{exp}).
The $g^\prime$ action on the KK modes is such that the invariant
states are
\begin{eqnarray}
&&\Phi^D_{m_1 m_2}+ (-1)^{m_1} \Phi^D_{m_1,-m_2}
{\rm \,\,\,\,\,\,\,\,\,\,\,doublet},\\
&&\Phi^T_{m_1 m_2} -(-1)^{m_1} \Phi^T_{m_1,-m_2} 
{\rm \,\,\,\,\,\,\,\,\,\,\,triplet}.
\end{eqnarray}
The KK towers of surviving states are then given in
Fig.~\ref{kk2}: 
for $m_2\neq 0$ they are identical, i.e. for each mass level
with $m_2\neq 0$ there is a complete multiplet of $SU(5)$.
If $m_2=0$ the doublet $\Phi^D$ has KK tower $M^2=m_1^2/R_1^2$
with {\it even} $m_1$'s, the triplet $\Phi^T$, instead, has tower
 $M^2=m_1^2/R_1^2$ with  {\it odd} $m_1$'s.
One can check that given any bulk field in a multiplet of $SU(5)$
the KK expansion shows similar features.
In detail, there are complete multiplets of $SU(5)$ for each KK level
with $m_2\neq 0$, incomplete multiplets for $m_2=0$ levels.
Moreover, in the $m_2=0$ case, the combination of the field
content of any $m_1=$even level with any $m_1=$odd level gives
again a complete multiplet of the $SU(5)$ group.

Given this, we can argue about the differential 
running of the coupling constants.
The states with $m_2\neq 0$ are irrelevant, since they fill
complete multiplets of $SU(5)$, i.e. $b_{ij;\,m_1 m_2}=0$ for
$m_2\neq 0$.
About the states with $m_2=0$, we have that a submultiplet with
even $m_1$ and one with odd $m_1$ contribute with
opposite coefficient, since they combine into a complete $SU(5)$ multiplet,
whose contribution to the differential running is zero.
We have then $b_{ij;\,2m_1 0}=-b_{ij;\,2m_1+1 0}$.
Moreover, the absence of fixed points ensures the absence
of extra localized matter\footnote{Notice that, from a string
perspective, there is a $g^\prime$ twisted sector, but all the states
have mass larger then $M_0\sim m_s^2 R_1$, where
$m_s$ is the string scale. Thus, these states are all heavier
than the string scale (cutoff) provided  that $R_1^{-1}<m_s$.}, 
so that the only contribution to
$b_{ij;\,0 0}$ comes from the expansion of six-dimensional fields.
This implies that the relation $b_{ij;\,2m_1 0}=-b_{ij;\,2m_1+1 0}$
holds also for $m_1=0$.
Then  Eq.~(\ref{run}) reduces to
\begin{equation}
\alpha_{ij}=b_{ij;\,00} \log\frac{R_1^{-1}}{M_Z}+ 
b_{ij;\,00} f\left[\frac{\Lambda}{R_1^{-1}}\right]
\end{equation}
with
\begin{equation}
\label{resto}
f\left[N\right]=
\left(\log N +
2\sum_{n=1}^{N}(-1)^n \log\frac{N}{n} \right).
\end{equation}
The function $|f[x]|$ is well defined for $x\rightarrow \infty$ and
converges in that limit to $\log(2/\pi)$.
In other terms,
the differential running due to the massless modes is completely
canceled by the threshold corrections arising at the compactification
scale $1/R_1$, that can be identified with the unification scale.
Thus, we conclude that in the described model the gauge symmetry breaking
has a breaking (unification) scale $M_{GUT}=1/R_1$. 

Unfortunately,
the described model cannot be relevant in model building, since
the orbifold action on $z_2$  is just a translation, and this 
clashes with the requirement of a four-dimensional
chiral spectrum. Such a problem can be avoided by adding an extra
orbifold operator, as discussed in the following section.

\subsection{A $T^2/Z_2\times Z_2^\prime$ model with non-local gauge symmetry
breaking}
In order to have a chiral 4d spectrum and non-local gauge symmetry breaking,
we can combine the $Z_2$ projection described in Sect.~\ref{roto} with the
$Z_2^\prime$ projection described in Sect.~\ref{trasl}. 
We have then a model with two orbifold projections, $g$ and $g^\prime$,
acting as
\begin{eqnarray}&&
g:z_1\rightarrow -z_1, \,\,\,\,\,\,\,\,\,\,\,\,\,\,\,\,\,\,\,\,  
g:z_2\rightarrow -z_2+\delta,\\&&
g^\prime:z_1\rightarrow z_1+\pi R_1,\,\,\,\,
g^\prime:z_2\rightarrow -z_2.
\end{eqnarray}
The requirement that $g$ and $g^\prime$ commute implies that $\delta=0$
or $\delta=\pi R_2$. 
In both cases the action of $g$ has fixed points in the internal space;
since we want a non-local gauge symmetry breaking we define 
$\gamma_g$ to be the identity matrix. In this way  $g$ does not break
the unified gauge symmetry and, moreover, it projects away  all the 
continuous Wilson lines.
In order to have a gauge symmetry breaking we define, 
$\gamma_g^\prime={\rm Diag}(-1,-1,-1,1,1)$, as done in Sect.~\ref{trasl}.

The orbifold group contains the operators $g$ and $g^\prime$, 
the first with local geometric action and not breaking the gauge 
symmetry, the second with non-local action and breaking the
gauge symmetry.
It also contains the mixed operator $g\cdot g^\prime$, breaking
the gauge symmetry. It has local action if $\delta=0$,
non-local action if $\delta=\pi R$.  In the latter case, the action
is non-local since it is a translation along $z_2$ (and the resulting
geometry is precisely the projective space described in\cite{Hebecker:2003we}).
This implies that only in the second case the gauge symmetry breaking
is completely non-locally realized. From this, we immediately conclude
that in the second case the  differential running of the coupling 
constants stops at a certain scale\cite{Hebecker:2003we}.
Still, it is interesting to check this via a direct computation,
in order to obtain a precise relation between the symmetry breaking scale
and the compactification scales.

\paragraph{The $\delta=0$ case: localized gauge symmetry breaking\\}
In the $\delta=0$ case the KK expansion of a multiplet in the 
fundamental representation of $SU(5)$ is given by the invariant
combinations
\begin{eqnarray}
&&\left(\Phi^D_{m_1,m_2}+\Phi^D_{-m_1,-m_2}\right)+
(-1)^{m_1}\left(\Phi^D_{m_1,-m_2}+\Phi^D_{-m_1,m_2 }\right),\\
&&\left(\Phi^T_{m_1,m_2}+\Phi^T_{-m_1,-m_2}\right)-
(-1)^{m_1} \left(\Phi^T_{m_1,-m_2}+\Phi^T_{-m_1,m_2 }\right).
\end{eqnarray}
This means that the towers for the doublet $\Phi^D$ and the triplet $\Phi^T$
coincide for $m_1,m_2\neq 0$, and so these states
are irrelevant for what concerns the differential running of the couplings.
For $m_1\ge 0,m_2=0$, instead, the two towers are identical but shifted: 
even positive $m_1$ states in the doublet, odd positive $m_1$ states in 
the triplet.
This is very similar to the situation studied in Sect.~\ref{trasl},
but with the crucial difference that in that case $m_1\in\mathbb Z$,
here the KK tower of invariant states has instead $m_1\in\mathbb N$.
Due to this, the running of the coupling constants receive threshold
corrections, but they cancel only {\it half} of the differential
running due to the zero modes (for a more direct geometric understanding
of this result see\cite{Hebecker:2002vm}).
Moreover, the remaining states with $m_1=0,m_2>0$, are present only 
in the doublet tower, making things even worse. Such a feature is
due to the fact that, if $\delta=0$, the breaking due to
$g\cdot g^\prime$ is localized not on isolated points in the internal 
space, but on the whole line $z_2=0$, $z_1\in  S^1$.
About the expansion of multiplets other then the fundamental, and the 
presence of localized extra matter, we can apply the same observation
of previous Sections, and conclude that the differential running of the
couplings will receive threshold corrections at the compactification scale,
but that, generically, they are able to only partially stop the differential 
running due to the zero modes\footnote{Clearly, given a {\it very 
specific choice} of the localized matter, it is in principle possible to 
stop the differential running of the couplings, at least in the case 
$R_2\lesssim 1/\Lambda$.}, and it is impossible to identify a breaking scale
$M_{GUT}$ such that Eq.~(\ref{run}) reduces to Eq.~(\ref{runc}).

\begin{figure}[t]
\begin{center}
\includegraphics[scale=0.5]{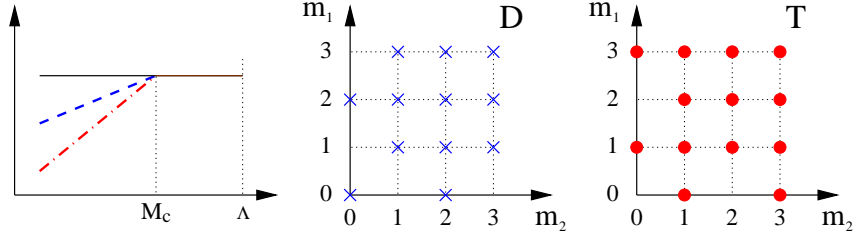}\vspace{-.1cm}
\caption{\footnotesize \it 
Kaluza-Klein expansion of a fundamental representation 
of $SU(5)$ in the $T^2/Z_2$ orbifold. 
The figure in the middle shows the KK tower of the doublet (D), 
a blue cross in a point with coordinates $m_1,m_2$
indicates the presence in the tower of a single state
with KK mass  $m^2=m_1^2/R_1^2+ m_1^2/R_2^2$. 
The figure on the right shows the tower for the triplet (T).
A similar expansion holds for any $SU(5)$ multiplets, this
implies that the running of the difference of the inverse
coupling constants is as given in the picture on the left,
i.e. it precisely stops, due to the KK contributions, at
the compactification scale $M_C=1/\sqrt{R_1 R_2}$.
In the model the breaking scale is then reproduced and explained
provided $M_C=M_{GUT}=3\times 10^{16} {\rm GeV}$.
}\label{kk3}
\end{center}
\end{figure}

\paragraph{The $\delta=\pi R_2$ case: non-local gauge symmetry breaking\\}
If $\delta=\pi R_2$, then the invariant combinations are
\begin{eqnarray}
&&\left(\Phi^D_{m_1,m_2}+(-1)^{m_2}\Phi^D_{-m_1,-m_2}\right)+(-1)^{m_1} 
\left(\Phi^D_{m_1,-m_2}+(-1)^{m_2}\Phi^D_{-m_1,m_2 }\right),\\
&&\left(\Phi^T_{m_1,m_2}+(-1)^{m_2}\Phi^T_{-m_1,-m_2}\right)-(-1)^{m_1} 
\left(\Phi^T_{m_1,-m_2}+(-1)^{m_2}\Phi^T_{-m_1,m_2 }\right),
\end{eqnarray}
for the doublet $\Phi^D$ and the triplet $\Phi^T$ respectively.
As shown in Fig.~\ref{kk3}, for $m_1,m_2\neq 0$ we have no difference
between the towers.
For $m_1=0$ the doublet tower contains only even positive $m_2$
states, while the triplet tower contains only odd positive $m_2$ states, 
and similarly for $m_2=0$: the doublet tower contains only even $m_1$'s
and the triplet only odd $m_1$'s. An identical expansion is present for
any bulk multiplet. Combining this with the  absence of extra localized
matter not filling $SU(5)$ multiplets\footnote{Indeed, all the singularities
of the orbifold are due to orbifold rotations respecting the $SU(5)$ symmetry,
and so the localized extra matter must fill complete $SU(5)$ multiplets.}, 
we conclude that the coefficient $b_{ij;\,m_1 m_2}$ in Eq.~(\ref{run})
are zero if both
$m_1$ and $m_2$ are non-zero. If $m_1$ is zero
$b_{ij;\,0\,2m_2}=-b_{ij;\,0\,2m_2+1}$. Similarly, if 
$m_2$ is zero,
$b_{ij;\,2m_1\,0}=-b_{ij;\,2m_1+1\,0}$.
Then Eq.~(\ref{run}) can be rewritten as
\begin{equation}
\alpha_{ij}=b_{ij;\,00} \log\frac{\sqrt{R_1^{-1} R_2^{-1}}}{M_Z}+ 
\frac{1}{2} b_{ij;\,00} f\left[\frac{\Lambda}{R_1^{-1}}\right]+
\frac{1}{2} b_{ij;\,00} f\left[\frac{\Lambda}{R_2^{-1}}\right],
\end{equation}
where the function $f$ has been defined in Eq.~(\ref{resto}).
Given the properties of $f$ we deduce the existence of a unification (breaking)
scale $M_{GUT}=\sqrt{R_1^{-1} R_2^{-1}}$.

\subsubsection{Generalization}
The explicit extension of the described computation to a generic orbifold is
unfortunately not available. Still, from the purely geometric arguments given
above, we can argue that in any model where the gauge symmetry breaking is
non-local, i.e. all the symmetry breaking orbifold
operators act as rototranslations, the differential running of the coupling
constants stops at a scale set by the compactification moduli.
In particular, in presence of $q$ orbifold operators responsible for
the gauge symmetry breaking, non-locality can be invoked provided
that each of them acts as a translation along some internal direction.
The scale of breaking is then given by the volume $V$ of the submanifold
spanned by such directions: $M_{GUT}=V^{-1/q}$.

\subsubsection{Non-local breaking vs. discrete/continuous Wilson lines}
We remark that the mechanism is peculiarly different from a discrete Wilson
line breaking. Consider, as an example,  a $T^2/Z_2\times Z_2^\prime$
orbifold with orbifold operators
\begin{eqnarray}
&& g: z_1\rightarrow -z_1, 
\,\,\,\,\,\,\,\,\,\,\,\,\,\,\,\,\,\,\,\, g:z_2\rightarrow -z_2;\\
&& g^\prime: z_1\rightarrow z_1+\pi R_1,\,\,\,\, g^\prime:z_2\rightarrow z_2.
\end{eqnarray}
The operator $g$ generates the orbifold $T^2/Z_2$, with four fixed points
$(z_1,z_2)=$ $(0,0)$, $(\pi R_1,0)$, $(0,\pi R_2)$, $(\pi R_1,\pi R_2)$,
where the breaking due to $g$ is localized.
The operator $g^\prime$ introduces a discrete Wilson line along $z_1$.
In other words, the fundamental space of $T^2/Z_2\times Z_2^\prime$
is the same as the one of $T^2/Z_2$  (with half volume) but the four
singularities are now not equivalent: there are two fixed points
$(z_1,z_2)=$ $(0,0)$, $(0,\pi R_2)$ where the gauge symmetry breaking due
to $g$ is localized, and two fixed points $(z_1,z_2)=$ $(\pi R^\prime_1,0)$,
$(\pi R^\prime_1,\pi R_2)$ where the breaking due to $g\cdot g^\prime$
is localized (we define $R^\prime_1=  R_1/2$).
This immediately clarifies that in such an orbifold the breaking is
always localized in points in the internal space: if $g$ breaks the gauge
symmetry then such a breaking is localized in the first two fixed points,
if instead $g\cdot g^\prime$ is responsible for the breaking, then it is
localized in the second fixed points.
If neither $g$ nor $g\cdot g^\prime$ break the gauge symmetry, then no gauge
symmetry breaking at all is present.

Given this, we conclude that a discrete Wilson line does not realize,
in general, a non-local gauge symmetry breaking, and we cannot expect the
presence of a scale of breaking.

This does not mean that {\it any} discrete Wilson line induces a
breaking without a scale: in some {\it very specific constructions}
such a feature can be achieved\footnote{This does not change our conclusion
about the qualitative difference between discrete Wilson lines and non-local
symmetry breaking: only in the latter the existence of a breaking scale is a 
{\it generic} feature.}. As a non-trivial example of this, in the last
section we show an $S^1/Z_2\times Z_2^\prime$ orbifold with a rank reducing
orbifold operator $g$ and a discrete Wilson line $g^\prime$ in which the
breaking due to $g^\prime$ has a scale of breaking, precisely given by the
compactification radius.

About the possibility that the described non-local gauge symmetry breaking 
may be completely reabsorbed into continuous Wilson lines, we remark that 
this is not the case in a semirealistic model with $\mathcal N=1$ SUSY
in 4d. 
Indeed, the SUSY condition implies that for each internal direction $z^m$
there is an orbifold operator $g_{z^m}$ acting as a rotation on it.
Since $g_{z^m}$ respects the SM gauge group, there are Wilson lines
$A^m$ only outside the SM generators, i.e. outside the Cartan subalgebra
of $SU(5)$. This implies that there can be continuous Wilson lines, but
their gauge symmetry breaking always induce a rank reduction.

\section{String models with non-local gauge symmetry breaking}
We pass to the study of backgrounds with six internal dimensions,
relevant in string model building: 
$T^2_1\times T^2_2\times T^2_3/Z_M\times Z^\prime_N$ orbifolds.
The orbifold groups are generated by $g$ and $g^\prime$, and are such that
both in heterotic and in Type I string case the gauge symmetry
breaking due to $g^\prime$ is non-local. Indeed, the details of the
breaking in these cases are exactly as in the field theory examples
studied in previous section. This is due to the fact that, in heterotic
and Type I string, the gauge degrees of freedom propagate in the whole
10d spacetime.
We comment in a separate subsection the gauge symmetry breaking that
occurs, instead, in the case of Type II string with Dp-branes ($p<9$).
In all the cases the resulting four dimensional models have
$\mathcal N=1$ SUSY.
This is the minimal choice with the described features,
since non-locality and $\mathcal N=1$ SUSY cannot be simultaneously obtained 
in $T^2_1\times T^2_2\times T^2_3/Z_M$ models.

\begin{figure}[t]
\begin{center}
\includegraphics[scale=0.5]{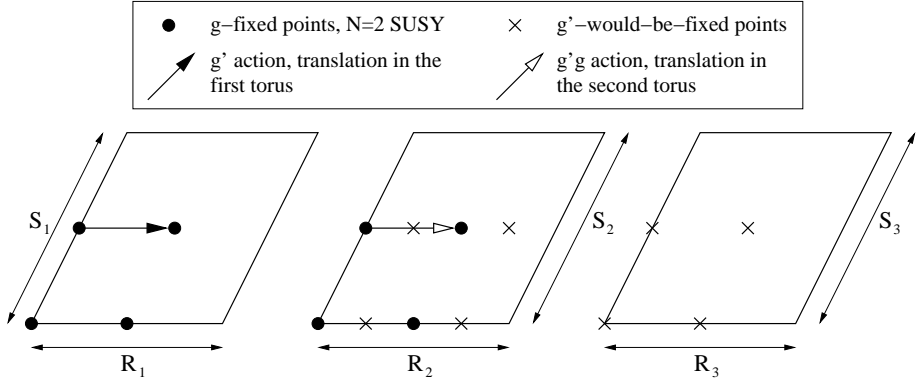}\vspace{-.1cm}
\caption{\small \it Internal geometry of the
$Z_{2}\times Z_{2}^{\prime}$ model. The action of 
$g$ is not free, the dots show its fixed points in
the first and second torus; since $g$ has no action
in the third torus each point $z_{3}\in\mathbb{C}$
is a (4d) fixed point preserving $\mathcal{N}=2$ SUSY.
The arrows show the action of $g^{\prime}$ ($g\,g^{\prime}$)
in the first (second) torus: a pure translation. 
The crosses show the would-be-fixed points of $g^{\prime}$.
}\label{classz2z2}
\end{center}
\end{figure}

We parameterize each $T^2_i$ torus with a complex variable $z_i$, with
periodicities $z_i\sim z_i + R_i$, $z_i\sim z_i + \tau_i R_i$. 
We allow non trivial complex structures $\tau_i$, previously set
to $i R_1/R_2$ (rectangular torus).
The orbifold group $Z_M$ is generated by $g$, a pure rotation acting on
each complex variable as $g:z_i\rightarrow e^{2\pi v_i/M} z_i$,
$v_i\in\mathbb Z$.
The orbifold group $Z^\prime_N$ is generated by $g^\prime$ acting as 
$g^\prime : z_i\rightarrow e^{2\pi v^\prime_i/N} z_i+\delta_i/N$,
$v^\prime_i\in\mathbb Z$, $\delta_i=a_i+b_i\tau_i$, $a_i,\,b_i\in\mathbb Z$.
We fix  $v^\prime_1=0$, $\delta_i\neq 0$ so that $g^\prime$ is a pure 
translation along $z_1$.
Since $g$ has local action and $g^\prime$ non local action, the gauge
symmetry breaking due to $g$ is localized, the one due to $g^\prime$ is
instead non-local.
Since the orbifold group contains also the mixed orbifold operators 
$g^n\cdot {g^\prime}^m$, the breaking due to $g^\prime$ is really
non-local only provided that all such operators (with $m\neq 0$)
act as rototranslations in the internal space.
This imposes strong constraints on $v$, $v^\prime$ and $\delta$,
to be combined with those on  $v$ and $v^\prime$  ensuring 
$\mathcal N=1$ SUSY in 4d, and with those on $\delta$ 
ensuring that the orbifold group is Abelian.
Examples of orbifolds fulfilling all these conditions are
\begin{itemize}
\item{$\bf T^6/Z_2\times Z_2^\prime$\\}
In this case $v=(1,1,0)$, $v^\prime=(0,1,1)$, $\delta=(1,1,0)$\footnote{It
is easy to check that the most general translation vector
$\delta_i=(a_i+b_i\tau_i)/2$ fulfills all the conditions provided that
$\delta_1\neq 0$ and at least one of the other two entries is non-zero.}.
There are two gauge symmetry breaking orbifold operators, $g^\prime$
and $g g^\prime$, both acting as rototranslations. The symmetry
breaking scale is $M_{GUT}=(R_1 R_2)^{-1/2}$.
The described geometry is shown in Fig.~\ref{classz2z2}.\vspace{6pt}
\item{$\bf T^6/Z_4\times Z_2^\prime$\\}
In this case $v=(1,1,-2)$, $v^\prime=(0,1,1)$, 
$\delta=(1+\tau_1,1+\tau_2,1/2)$ with $\tau_1=\tau_2=i$.
There are four gauge symmetry breaking orbifold operators,
$g^n g^\prime$ for $n=0,\dots,3$, all acting as rototranslations.
The symmetry breaking scale is $M_{GUT}=(R_1 R_2 R_3)^{-1/3}$.\vspace{6pt}
\item{$\bf T^6/Z_3\times Z_3^\prime$\\}
In this case $v=(1,1,0)$,
$v^\prime=(0,1,1)$, $\delta_i=1+\tau_i$, $\tau_i=e^{2\pi i/3}$.
All the operators $g^n {g^\prime}^m$ with $m\neq 0$ break the
GUT symmetry and act as rototranslations.
The symmetry breaking scale is $M_{GUT}=(R_1 R_2 R_3)^{-1/3}$.
\end{itemize}

\subsection{Open strings on freely acting orbifolds:
internal vs external orbifold action}
In type II string theory the gauge degrees of freedom propagate on
Dp-branes, hypersurfaces of $p+1$-dimensionality.
In case $p<9$, then, the orbifold action can be {\it internal}
to a single stack of D-branes, i.e. it  can map it to itself, but it can
also be {\it external}, i.e. it can map a stack of D-branes into a
different one, localized at a different point in the internal space.
In the first case the physics is as described in the field theory
examples given in previous section.
In the second case there is a rank reduction of the gauge group:
if an orbifold action is external to a stack of D-branes, then it
maps it into a different stack, identifying the corresponding gauge
groups.
As an example, consider a stack $A$ of D-branes with $U(5)_A$
gauge group, a stack $B$ of D-branes with $U(5)_B$ gauge
group and an orbifold operator $g$ mapping $A$ in $B$\footnote{
Notice that the two stacks must be identical, otherwise
the orbifold action identifying $A$ with $B$ cannot be a symmetry of
the system.}.
Whatever the action of $g$ is, the unbroken gauge symmetry is $U(5)$, 
combination of $U(5)_A$ with $U(5)_B$.
Notice that, if $g$ is a rotation, we can continuously move
the $A$ and $B$ stacks till they coincide, in a fixed point of $g$,
a gauge enhancement occurs and the gauge symmetry breaking
due to $g$
is a standard rank-preserving symmetry-breaking of the enhanced
$U(10)$ gauge group (notice that the surviving gauge group is also
an enhancement of $U(5)$). 
In this sense, in the purely rotational case, the
picture is completely dual to the $p=9$ case, with the displacement
between $A$ and $B$ stacks replaced by a continuous rank-reducing
Wilson line.
If instead the action of $g$ is a translation, then the distance between
$A$ and $B$ is fixed and no gauge enhancement is possible\footnote{
In other words, in the dual picture, the corresponding continuous
rank-reducing Wilson line is projected out of the spectrum by $g$,
but $g$ itself can be seen as a sort of discrete rank-reducing Wilson line.}.

From the description above we conclude that an orbifold
model with only {\it internal} orbifold actions can be completely
described as in the field theory examples given in previous sections.
If instead the orbifold actions are completely {\it external}, then
the gauge symmetry breaking always includes a rank reduction, and
a breaking of the kind $SU(5)\rightarrow SU(3)\times SU(2)\times U(1)$
is never possible.
The case with {\it both} internal {\it and} external orbifold actions,
instead, is more intriguing, and we show the possibilities that it offers
in the following subsection, with an explicit example.

Notice that the description of the gauge symmetry breaking in presence
of internal or external translational orbifold actions 
(also called parallel and perpendicular actions respectively) is very
close to the SUSY breaking one (as an example, see\cite{Antoniadis:1999ux}).

\begin{figure}[t]
\begin{center}
\includegraphics[scale=0.5]{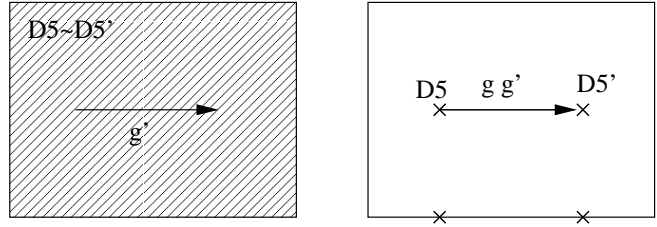}\vspace{-.3cm}
\caption{\small \it
The background described in Fig.~\ref{classz2z2} with two stacks of
5 D5-branes filling the first torus and located in two would-be-fixed
points of $g^{\prime}$ in the second torus (the third torus is not shown,
the D5-branes all sit in $g^\prime$-fixed points).
}\label{z2z2seminonlocal}
\end{center}
\end{figure}

\subsubsection{A $Z_{2}\times Z_{2}^{\prime}$ model with mixed 
{\it internal} and {\it external} orbifold action: localized
breaking with a breaking scale}
The combination of internal and external orbifold actions is peculiar,
since in this case a discrete Wilson lines is naturally combined
with a rank-reducing orbifold projection.
As an example, we consider the $Z_{2}\times Z_{2}^{\prime}$ geometry of
Fig.~\ref{classz2z2} and embed two stacks of 5 D5-branes, filling the first
torus and localized into two different would-be-fixed points of $g^{\prime}$
in the second torus, as shown in Fig.~\ref{z2z2seminonlocal}.
The gauge group, in absence of any orbifold projection, is
$U(5)_{D5}\times U(5)_{D5^{\prime}}$.
The action of $g^{\prime}$ is free and internal, since it maps each stack of
D-branes into itself. We can embed it into the gauge degrees of freedom
such that $g^\prime:U(5)\rightarrow U(3)\times U(2)$ both for D5 and
D$5^\prime$ -branes.
The action of $g$ is instead non-free and external, since it maps one
stack of D-branes into the other in the second torus, so that 
$g:U(5)_{D5}\times U(5)_{D5^{\prime}}\rightarrow U(5)_g$.
Similarly, the action of $g g^{\prime}$ is also external, and induces
an identification between the two $U(5)$'s, but the breaking is
$gg^\prime:U(5)_{D5}\times U(5)_{D5^{\prime}}\rightarrow U(5)_{g^\prime}$,
such that $U(5)_g\cap U(5)_{g^\prime}=U(3)\times U(2)$, so that the
ending unbroken group is then just the latter group.

\begin{figure}[t]
\begin{center}
\includegraphics[scale=0.6]{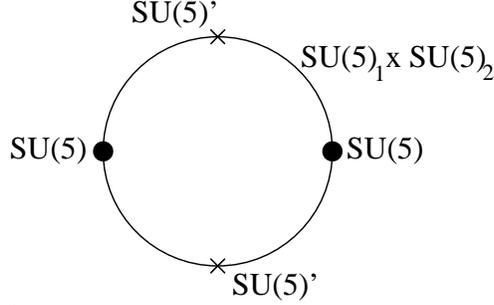}\vspace{-.3cm}
\caption{\small \it
An $S^{1}/Z_{2}\times Z_{2}^{\prime}$ field theory model.
The bulk symmetry is $SU(5)_{1}\times SU(5)_{2}$, broken to
$SU(5)$ and to $SU(5)^{\prime}$ in the $g$ and $g\,g^{\prime}$
fixed points (dots and crosses respectively).
The surviving gauge group is just the SM gauge group:
$SU(5)\cap SU(5)^{\prime}=SU(3)\times SU(2)\times U(1)$.
 }\label{esseuno}
\end{center}
\end{figure}

It is interesting to reproduce these features in a simplified field theory
model, with a single extra dimension parameterized by $x\sim x+2\pi R$
and compactified on $S^{1}/Z_{2}\times Z_{2}^{\prime}$\footnote{It should be
possible to obtain such a field theory model, with $SU(5)$ groups rather then
$U(5)$ groups, from the described open string model, by shrinking the
second and third torus, and the vertical direction of the first torus, and neglecting
the corresponding KK states.}.
The action of the $Z_{2}$ operator is $g:\,x\rightarrow -x$, with fixed
points $0$ and $\pi R$, while the action of the $Z_{2}^{\prime}$ operator is
$g^{\prime}:\,x\rightarrow x+\pi R$.
Thus, the mixed operator $g g^\prime$ has fixed points $\pi R/2$ and $3\pi R/2$
(see Fig.~\ref{esseuno}) and $g^\prime$ can be seen as a discrete Wilson line.
The bulk gauge symmetry is $SU(5)_{1}\times SU(5)_{2}$, $g$ acts on
the gauge group such that $g:SU(5)_{1}\leftrightarrow SU(5)_{2}$.
More precisely, defining $T^{ab}_{i}$ a generator of $SU(5)_{i}$\footnote{
We take $a,\,b=1,2\dots,5$, and define $T^{aa}$ as the 5 Cartan generators of
$U(5)$, from which we exclude the ``diagonal''
generator $\sum_{a} T^{aa}$.} the identification is
$g:T^{ab}_{1}\leftrightarrow T^{ab}_{2}$, and the surviving gauge group in
$0$ and $\pi R$ is a {\it diagonal} $SU(5)$ generated by
$T^{ab}_{1}+ T^{ab}_{2}$.
We embed then $Z_{2}^{\prime}$ action as 
$g^{\prime}:T^{ab}_{1}\rightarrow \delta^{a}_{c}T^{cd}_{2}\delta^{b}_{d}$
with $\delta$ a diagonal matrix: $\delta={\rm Diag[-1,-1,-1,1,1]}$.
In this way the surviving gauge group in $\pi R/2$ and $3\pi R/2$ is
the one left invariant by the operator $g g^\prime$ and it is a 
{\it different} $SU(5)^{\prime}$, such that the intersection of the
two gauge  groups is just $SU(3)\times SU(2)\times U(1)$, the 
Standard Model (SM) gauge group. 
The breaking is local, since the $SU(5)$ symmetry preserved in $0$ is
generically broken in $\pi R/2$, but the differential running of the
coupling constants is generated only by the bulk degrees of freedom, and it
stops precisely at the scale $M_{GUT}=R^{-1}$.
There is no fixed-point-contribution to the differential running since only
full multiplets of $SU(5)$ ($SU(5)^{\prime}$) can be localized there,
and the SM gauge group is embedded exactly in the {\it same} way in
$SU(5)$ and in $SU(5)^{\prime}$.
The last point is crucial: a multiplet of $SU(5)$ behaves as a multiplet of
$SU(5)^{\prime}$ 
under the action of the SM gauge group, and they both contribute universally to
the running of the coupling constants.
We expect the same argument to be valid for any localized contribution to the
action.
Thus, in this model there is a scale of breaking, given by the length
of the unique compact dimension $M_{GUT}=R^{-1}$.

The described mechanism can be embedded, in principle, also in
an heterotic string theory context, in presence of rank-reducing
continuous Wilson lines. We can indeed have an orbifold model with
$U(10)$ gauge symmetry in the bulk (the breaking
$SO(32)\rightarrow U(10)\times \mathcal G_{Hidden}$ can be
realized via discrete Wilson lines), and two orbifold operators $g$ and
$g^\prime$.
We take $g$ such that it breaks $U(10)$ to $U(5)\times U(5)$, and
$g^\prime$, freely acting, such that each $U(5)$ is broken to
$U(3)\times U(2)$. We can choose $g$ and $g^\prime$ such that any
combination $g^n {g^\prime}^m$ breaks $U(10)$ to some
$U(5)^\prime\times U(5)^\prime$. In this way all the singularities in
the orbifold respect some $U(5)$ symmetry, and the scale of breaking
$U(5)\times U(5)\rightarrow U(3)\times U(2)\times U(3)\times U(2)$
is set by a {\it single} internal dimension $R_{g^\prime}$.
By adding a continuous Wilson line we can then identify the two $U(5)$
factors, and have a model with a massless $U(3)\times U(2)$ gauge
group, with coupling unification at $1/R_{g^\prime}$, and an
extra twin gauge group $U(3)^\prime\times U(2)^\prime$, broken by
the continuous Wilson line.
Notice also that such a model can arise only in $SO(32)$ heterotic
string, where examples of orbifold models with $U(10)$ (or $SO(20)$)
gauge groups are present\cite{me}.

\end{document}